\documentstyle[aps,twocolumn]{revtex}

\input{epsfig.sty}

\def \no{\noindent}
\def \be{\begin{eqnarray}}
\def \ee{\end{eqnarray}}

\begin{document}
\twocolumn[\hsize\textwidth\columnwidth\hsize\csname @twocolumnfalse\endcsname
\title{ EVOLUTION OF PROTO--NEUTRON STARS WITH QUARKS}

\author{Jos\'e A. Pons, Andrew W. Steiner, Madappa Prakash, 
and James M. Lattimer}
\address{Department of Physics and Astronomy, SUNY at Stony Brook, Stony Brook,
NY 11794-3800}

\date{\today}

\maketitle

\begin{abstract}
Neutrino fluxes from proto-neutron stars with and without quarks are
studied.  Observable differences become apparent after 10--20 s of
evolution.  Sufficiently massive stars containing negatively-charged,
strongly interacting, particles collapse to black holes during the
first minute of evolution.  Since the neutrino flux vanishes when a
black hole forms, this is the most obvious signal that quarks (or
other types of strange matter) have appeared.  The metastability
timescales for stars with quarks are intermediate between those
containing hyperons and kaon condensates. \\

\no{PACS: 97.60.Jd, 21.65.+f, 12.39.-x, 26.60.+c}   \\
\end{abstract}
]


A proto-neutron star (PNS) is born following the gravitational
collapse of the core of a massive star, in conjunction with a
successful supernova explosion.  During the first tens of seconds of
evolution, nearly all ($\sim$ 99\%) of the remnant's binding energy is
radiated away in neutrinos of all flavor \cite{BL86}.  The
$\nu-$luminosities and the evolutionary timescale are controlled by
several factors, such as the total mass of the PNS and the
$\nu-$opacity at supranuclear density, which depends on the
composition and equation of state (EOS).  One of the chief objectives
in modeling PNSs is to infer their internal compositions from
$\nu-$signals detected from future supernovae like SuperK, SNO and
others under consideration, including UNO \cite{UNO}.

In their landmark paper, Collins and Perry \cite{CP75} noted that the
superdense matter in neutron star cores might consist of weakly
interacting quarks rather than of hadrons, due to the asymptotic
freedom of QCD.  The appearance of quarks causes a softening
of the EOS which leads to a reduction of the maximum mass and radius
\cite{LP01}.  In addition, quarks would alter $\nu-$emissivities and
thereby influence the surface temperature of a neutron star
\cite{PPLS00} during the hundreds of thousands or millions of years
that they might remain observable with such instruments as HST,
Chandra and XMM.  Quarks would also alter the spin-down rates of
neutron stars \cite{GPW97}.

Many calculations of dense matter predict the appearance of other
kinds of exotic matter in addition to quarks: for example, hyperons or
a Bose (pion, kaon) condensate (cf. \cite{Pra97} and references therein).
An important question is whether or not $\nu$ observations from a
supernova could reveal the presence of such exotic matter, and further
could unambiguously point to the appearance of quarks.  The detection
of quarks in neutron stars would go a long way toward the delineation
of QCD at finite baryon density which would be complementary to
current Relativistic Heavy Ion Collider experiments, which
largely address the finite temperature, but baryon-poor regime.  

An important consequence of the existence of exotic matter in neutron
stars (in whatever form, as long as it contains a negatively charged
component), is that a sufficiently massive PNS becomes metastable
\cite{Metas}.  After a delay of up to 100 s, depending upon
which component appears, a metastable PNS collapses into a black hole
\cite{kj,PRPLM99,Pons01}.  The collapse to a black hole proceeds on a
free-fall timescale of less than a millisecond \cite{BST96}, much shorter 
than $\nu-$diffusion times, and the neutrinos still trapped in the inner
regions cannot escape.  Such an event should be straightforward to
observe as an abrupt cessation of $\nu-$flux when the instability is
triggered \cite{Bur88}.  

The evolution of the PNS in the so-called Kelvin-Helmholtz phase,
during which the remnant changes from a hot, $\nu-$trapped, and
lepton-rich object to a cold and $\nu-$free star, occurs in
near-hydrostatic equilibrium.  The $\nu-$matter interaction timescales
are much smaller than the dynamical timescale of PNS evolution,
which is of the order of seconds. Thus, until neutrinos enter the
semi-transparent region at the edge of the star, they remain close to
thermal equilibrium with matter, and may be treated in the diffusion
approximation.

In this Letter we provide a benchmark calculation with quarks by
solving the general relativistic $\nu-$transport and hydrostatic
equations (as in \cite{PRPLM99}), and then compare our results with
those of our previous work \cite{PRPLM99,Pons01} in which other
compositions were studied.  In addition, we assess the prospects of
observing PNS metastability and its subsequent collapse to a black
hole through observations in current and planned detectors.

The essential microphysical ingredients in our study are the EOS of
dense matter and its associated $\nu-$opacity.  We begin by
considering two generic compositions: charge-neutral, beta
equilibrated matter containing (i) nucleons only ($np$) and (ii)
nucleons with quark matter ($npQ$).  In the $npQ$ case, a mixed phase
of baryons and quarks (pure quark matter exists only for very large
baryon densities, except for extreme choices of model parameters) is
constructed by satisfying Gibbs' phase rules for mechanical, chemical
and thermal equilibrium \cite{G92}.  The EOS of baryonic matter is
calculated using a field-theoretic model at the mean field level
\cite{MS96}.  The results reported with this EOS are quite general, as
we verified by alternatively using a potential model approach
\cite{Pra97}. The quark matter EOS is calculated using a MIT
bag-like model (similar results are obtained with the
Nambu--Jona-Lasinio quark model).  The details of the EOS may be found
in Ref.~\cite{SPL00}.  We use $\nu-$opacities\cite{RPL98,SPL01}
consistent with the EOS.  When quarks appear, the $\nu$
absorption and scattering cross sections dramatically decrease, the
precise reduction being sensitive to the thermodynamic conditions in
the mixed phase \cite{SPL01}.


\begin{figure}[ht]
\centerline{{\epsfxsize=4.1in \epsfbox{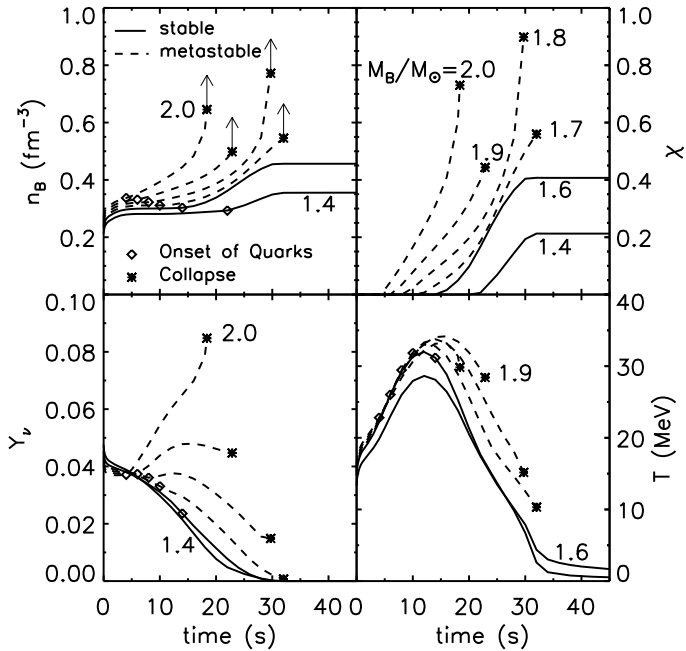}} }
\caption{Evolutions of the central baryon density $n_B$, $\nu$ 
concentration $Y_\nu$,
quark volume fraction $\chi$ and temperature $T$ for different 
baryonic masses $M_B$.
Solid lines correspond to stable stars; stars with larger masses are
metastable (dashed lines).  Diamonds indicate when quarks appear at the
star's center, and asterisks denote when metastable stars become
gravitationally unstable.}
\label{evol1} \end{figure}


Fig. \ref{evol1} shows the evolutions of some thermodynamic quantites
at the center of $npQ$ stars of various, fixed, baryonic masses
($M_B$).  In the absence of accretion, $M_B$ remains constant during
the evolution, while the gravitational mass $M_G$ decreases.  With the EOS
utilized, stars with $M_B\lesssim1.1$ M$_\odot$ do not contain quarks
and those with $M_B\gtrsim1.7$ M$_\odot$ are metastable.  The latter
value is $\sim0.05$ M$_\odot$ larger than the maximum mass for cold,
catalyzed $npQ$ matter, because the maximum mass of hot $\nu-$free
$npQ$ matter is this much less than that of hot $\nu-$trapped
matter\cite{SPL01}.  Generally, due to the high lepton ($\nu$) content
initially present in the PNS, the electron chemical potential at the
center is too large for quarks to exist. For sufficiently massive
stars, quarks eventually appear after a certain amount of $\nu$ loss
occurs.  For the $M_B=1.6$ M$_\odot$ star, for example, quarks appear
after about 15 s (indicated by a diamond).  Thereafter, the star's
central density increases for a further 15-20 s, until a new
stationary state with a quark-hadron mixed phase core is reached (for
stable stars) or for $M_B\gtrsim 1.7$ M$_\odot$, instability occurs
(indicated by asterisks).  It is interesting that for this EOS the
lifetimes for all masses are restricted to the range 10--30 s, and
slowly decrease with increasing mass.  The appearance of quarks is
accompanied by an increase in $Y_\nu$ because of the depletion of
electrons; for the largest masses, the increase is very large.
 

\begin{figure}[ht]
\centerline{{\epsfxsize=3.6in \epsfbox{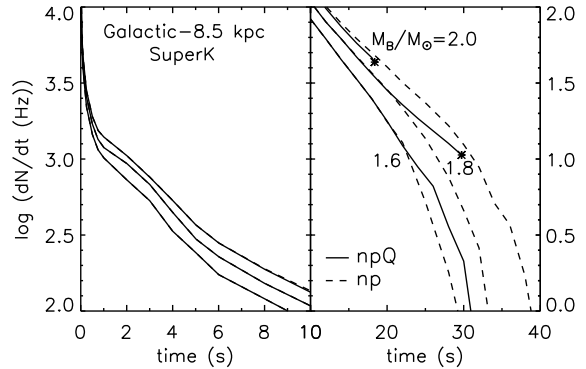}} }
\caption{A comparison of $\bar\nu_e$ count rates expected in SuperK
from a PNS containing either $np$ or $npQ$ matter.  The left panel
shows times less than 10 s, while the right panel shows times greater
than 10 s.}
\label{dndt}
\end{figure}

To point out the major differences one might observe between the $np$
and $npQ$ cases, we have estimated the $\bar\nu_e$ count rate in the
SuperK detector in Figure \ref{dndt}.  For this estimate, we assumed
the total $\nu-$luminosity from a PNS at 8.5 kpc distance,
corresponding to a Galactic supernova, was equally divided among the
six $\nu$ species.  The $\nu-$energy spectra were taken to be
Fermi-Dirac with zero chemical potential and a temperature
corresponding to the matter temperature in the PNS where the
$\nu-$optical depth was approximately unity.  Only the dominant
reaction, $\bar\nu_e$ absorption on protons, was included.  (For
details, see \cite{Pons01}.)  It is difficult to discern much
difference in the early ($t<10$ s) count rates from $np$ and $npQ$
stars. For stars with $M_B < 1.7$ M$_\odot$, this is because quarks
have not yet appeared.  For more massive stars, the fact that
neutrinos are strongly trapped inhibits any discriminatory signal from
reaching the surface before this time. The signals at later times
($t\gtrsim 25$ s), however, are substantially larger for the $npQ$
case, due to the decrease in $\nu-$opacity of $npQ$ matter and the
increased binding energy of $npQ$ stars.  Most importantly, the
$\nu-$signal from metastable $npQ$ stars halts abruptly when the
instability occurs.  Qualitatively, these features are also found for
$npH$ and $npK$ stars\cite{PRPLM99,Pons01}.

\begin{figure} 
\centerline{{\epsfxsize=3.2in \epsfbox{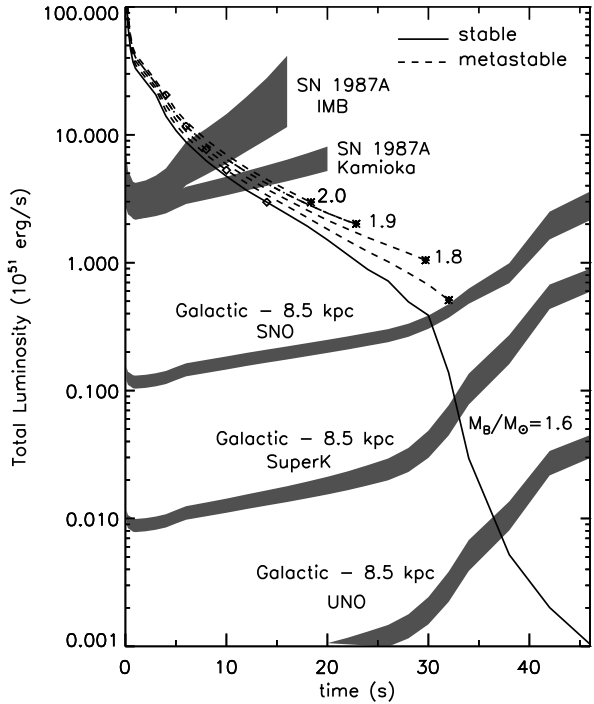}} }
\caption{The total $\nu-$luminosity for $npQ$ stars of various baryon
masses.  Shaded bands illustrate the limiting luminosities corresponding
to count rates of 0.2 Hz for the indicated supernovae in some detectors.} 
\label{lum1} 
\end{figure}

We compare $\nu-$signals observable with different detectors in Fig.
\ref{lum1}, which displays the $\nu-$light curves as a function of
$M_B$ for $npQ$ stars.  The two upper shaded bands correspond to
estimated SN 1987A (50 kpc distance) detection limits with KII and
IMB, and the lower bands correspond to estimated detection limits in
SNO, SuperK, and UNO, for a Galactic supernova (8.5 kpc distance).
The detection limits have been set to an $\bar\nu_e$ count rate
$dN/dt=0.2$ Hz\cite{Pons01} with the same assumptions as in
Fig. \ref{dndt}.  The general rise with time of the detector limits is
chiefly due to the steady decrease in the $\bar\nu_e$ average energy.
It is possible that these limits are too conservative and could be
lowered with identifiable backgrounds and knowledge of the direction
of the signal.  The width of the bands represents the uncertainty in
the $\bar\nu_e$ average energy due to the flux-limited diffusion
approximation.  We conclude that it should be possible to distinguish
between stable and metastable stars, since the luminosities when
metastability is reached are always above conservative detection
limits.


The drop in $\nu-$luminosity for stable stars is associated with the
end of the Kelvin-Helmholtz epoch when the PNS is becoming optically
thin.  This portion of the $\nu-$light curve is approximate due to the
breakdown of the diffusion approximation.  It is an apparent
coincidence that this occurs simultaneously with the collapse of the
lower mass metastable stars.

\begin{figure}[ht]
\centerline{{\epsfxsize=3.0in \epsfbox{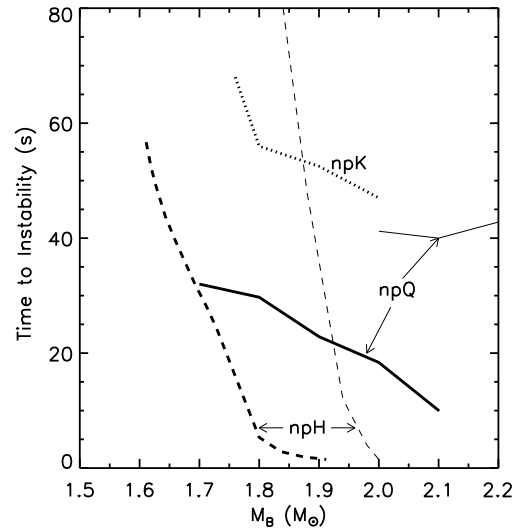}} }
\caption{Lifetimes of metastable stars versus the PNS $M_B$
for various assumed compositions.  Thick lines denote
cases in which the maximum masses of cold, catalyzed stars are near
$M_G\simeq1.45$ M$_\odot$, which minimizes the metastability
lifetimes.  The thin lines for the $npQ$ and $npH$ cases are for EOSs
with larger maximum masses ($M_G=1.85$ and 1.55 M$_\odot$,
respectively.}
\label{ttc}
\end{figure}

Our choice of bag constant, $B=150$ MeV fm$^{-3}$, in conjunction with
the baryonic EOS we used, was motivated to maximize the extent of the
quark matter phase in a cold neutron star, and was limited by the
necessity of producing a maximum mass cold star in line with accurate
observational constraints ($M_G=1.444$ M$_\odot$ \cite{tc}).
Increasing $B$, or employing an alternative quark EOS that otherwise
produces a larger maximum mass, delays the appearance of quarks and
raises the metastability window to larger stellar masses.
Necessarily, this results in an increased timescale for metastability
for a given mass, and a lower $\nu-$luminosity when metastability
occurs.  Fig. \ref{ttc} shows the relation between time to instability
and $M_B$ for the original case ($B=150$ MeV fm$^{-3}$, thick solid
curve) and a case with $B=200$ MeV fm$^{-3}$ (thin solid curve), in
which the maximum gravitational mass of a cold neutron star is about
1.85 M$_\odot$.  For the latter case, the metastability timescales lie
in a narrow range 40--45 s.  These, and the metastability masses, are
both larger than obtained for $B=150$ MeV fm$^{-3}$ and have narrower
ranges.  Further increases in $B$ diminish the size of the instability
window, because the appearence of quarks is shifted to progressively
larger densities.


Figure \ref{ttc} also compares the metastability time-$M_B$ relation
found for matter containing hyperons ($npH$, dashed lines\cite{PRPLM99})
or matter with kaons ($npK$, dotted line\cite{Pons01}) instead of
quarks.   All three types of strange matter are suppressed by trapped
neutrinos\cite{Pra97,SPL00}, but hyperons always exist in $npH$ matter
at finite temperatures and the transition to quark matter can occur at
lower densities than that for very optimistic kaon cases \cite{Pons01}. 
Thus, the metastability timescales for $npH$ matter can be very short,
and those for $npK$ matter are generally larger than for $npQ$ matter.  
Note the relatively steep dependence of the metastability time with
$M_B$ for $npH$ stars, which decreases to very small values near the
maximum mass limit of hot, lepton-rich, stars. The thick $npH$  and
$npQ$ lines, as well as the $npK$ line, represent minimum metastability
times for a given $M_B$ as discussed above.  The thin $npQ$ and $npH$
lines are for EOSs with larger cold, catalyzed maximum mass.  

Clearly, the observation of a single case of metastability, and the
determination of the metastability time alone, will not necessarily
permit one to distinguish among the various possibilities. Only if the
metastability time is less than 10--15 s, could one decide on this
basis that the star's composition was that of $npH$ matter.  However,
as in the case of SN 1987A, independent estimates of $M_B$ might be
available \cite{THM90BB95}.  In addition, the observation of two or
more metastable neutron stars might permit one to differentiate among
these models, but given the estimated rate of Galactic supernova (1
per 30--50 years \cite{snr}), this may prove time-consuming.
 
Our study has focused on times longer than approximately 1 s after
core bounce, after which effects of dynamics and accretion become
unimportant.  Studies of the $\nu$ signal during the first second,
during which approximately 1/3 of the energy is emitted, and at late
times, as the star becomes optically thin to neutrinos, requires more
accurate techniques for $\nu-$transport.  In addition, the earliest
time periods require the incorporation of hydrodynamics \cite{Trans}.

Our conclusions are that (1) the metastability and subsequent collapse
to a black hole of a PNS containing quark matter, or other types of
matter including hyperons or a Bose condensate, are observable in
current and planned $\nu$ detectors, and (2) discriminating among
these compositions may require more than one such observation.  This
highlights the need for breakthroughs in lattice simulations of QCD at
finite baryon density to unambiguously determine the EOS of
high density matter.  In the meantime, intriguing possible extensions
of PNS simulations with $npQ$ matter include the consideration of
heterogenoeus structures \cite{CGS00}, quark matter superfluidity
\cite{CR00} and coherent $\nu-$scattering on droplets \cite{RBP00}.

This work was supported by U.S. Department of Energy grants
 DOE/DE-FG02-87ER-40317 and
DOE/DE-FG02-88ER-40388. 
JAP acknowledges J.A. Miralles for interesting discussions.

{}

\end{document}